\documentclass[showpacs,twocolumn,prl,superscriptaddress,notitlepage]{revtex4-2}

\usepackage[x11names]{xcolor} 
\usepackage{graphicx}
\usepackage{tikz}
\usetikzlibrary{decorations.pathreplacing,decorations.pathmorphing}
\usepackage{rotating}
\usepackage{pstricks,pst-plot,pst-eps}
\usepackage{subfig}
\usepackage{braket}
\usepackage{tikz}
\usepackage{float}
\usepackage{color}
\usepackage{amsmath}
\usepackage{amsfonts}
\usepackage{siunitx}
\usepackage[ruled]{algorithm2e}
\usepackage{multirow}
\usepackage{braket}
\usepackage{url}
\usepackage{appendix}

\begin{document}

\title{Special Issue: Commemorating the 110th Anniversary of TANG Au-chin's Birthday \\ Calculation of the Green’s function on near-term quantum computers via Cartan decomposition}

\author{Lingyun Wan}
\affiliation{State Key Laboratory of Precision and Intelligent Chemistry, University of Science and Technology of China, Hefei, Anhui 230026, China}

\author{Jie Liu}
\email{liujie86@ustc.edu.cn}
\affiliation{Hefei National Laboratory, University of Science and Technology of China, Hefei 230088, China}

\author{Jinlong Yang}
\affiliation{State Key Laboratory of Precision and Intelligent Chemistry, University of Science and Technology of China, Hefei, Anhui 230026, China}
\affiliation{Hefei National Laboratory, University of Science and Technology of China, Hefei 230088, China}
  
\date{\today}

\begin{abstract}
Accurate computation of the Green’s function is crucial for connecting experimental observations to the underlying quantum states. A major challenge in evaluating the Green’s function in the time domain lies in the efficient simulation of quantum state evolution under a given Hamiltonian—a task that becomes exponentially complex for strongly correlated systems on classical computers. Quantum computing provides a promising pathway to overcome this barrier by enabling efficient simulation of quantum dynamics. However, for near-term quantum devices with limited coherence times and fidelity, the deep quantum circuits required to implement time-evolution operators present a significant challenge for practical applications. In this work, we introduce an efficient algorithm for computing Green's functions via Cartan decomposition, which requires only fixed-depth quantum circuits for arbitrarily long time simulations. Additionally, analytical gradients are formulated to accelerate the Cartan decomposition by leveraging a unitary transformation in a factorized form. The new algorithm is applied to simulate long-time Green’s functions for the Fermi-Hubbard and transverse-field Ising models, extracting the spectral functions through Fourier transformation.
\end{abstract}

\maketitle

\section{Introduction}
Quantum computing is poised to revolutionize various fields, including condensed matter physics and physical chemistry. A key application of quantum computers is the accurate prediction of quantum mechanical properties of strongly correlated systems~\cite{Fey82}, which are intractable for classical computers due to exponential computational complexity. With the rapid advancement of quantum computers, numerous quantum algorithms have been proposed for predicting ground- and excited-state eigenenergies and eigenstates~\cite{DuXuPen10,PerMcCSha14,MalBabKiv16,KanMezTem17,HemMaiRom18,AruAryBab20,CaoRomOls19,McAEndAsp20,ZhoWanDen20,CerArrBab21,LiuFanLi22,ShaSheFan22,HugGorRub22,RobMotHas25}. While, other important quantities, such as Green’s functions, have received relatively less attention in recent quantum algorithm developments. Among these, single-particle Green's functions are indispensable tools in both chemistry and physics for understanding the fundamental properties of many-body quantum systems. By encapsulating the correlation between particles in a system, single-particle Green's functions serve as a bridge between the microscopic quantum description and macroscopic observables, making them central to the study of both equilibrium and non-equilibrium processes across a range of systems~\cite{fetter2012quantum}.

Numerous numerical methods have been suggested for the approximate computation of the fully interacting Green's function~\cite{WilFeiLan82,KotSavHau06,HamesJoh16}. One leading strategy is based on the Lehmann representation of the spectral function, which requires access to the exact eigenstates and eigenenergies of the Hamiltonian. In the context of quantum computing, the subspace search variational quantum eigensolver (VQE)~\cite{NakMitFuj19,EndKurNak20}, the
multistate contracted VQE~\cite{ParHohMcM19}, and quantum equation of motion~\cite{RizLibTac22,OllKanChe20} have been suggested to compute excited states and the transition amplitude. However, in order to achieve the required accuracy, it is generally unclear how many excited states are necessary. To avoid the sum over excited states, the continued fraction representation of the Green’s function can be obtained through a quantum version of Lanczos recursion~\cite{Bak21,JamAgaLup21} or one can employ an auxiliary state generated from variational quantum response theory to compute the Green's function~\cite{CaiFanFan20,HuaCaiLi22,CheNusTil21,KokLabFre24}. 

Another promising approach is to leverage quantum dynamical simulation (QDS) algorithms to simulate the real-time evolution of quantum states, which lies at the heart of computing the Green’s function in the time domain. Given an initial state $\ket{\Psi_0}$, its time evolution under a given Hamiltonian $\mathcal{H}$ is described as
\begin{equation}
   \ket{\Psi(t)} = e^{-i\mathcal{H}t} \ket{\Psi_0},
\end{equation}
which can be implemented with Suzuki-Trotter expansion on a quantum computer~\cite{berry2007efficient,RogCar19,KosMat20}. In contrast, Taylor expansion-based Hamiltonian simulation approach has a better asymptotic scaling, in which the time evolution operator is approximated as a linear combination of unitary operators~\cite{BerChiCle15}. On a fault-tolerant quantum computer, these methods provide a simple and accurate scheme for computing the Green's function but at the expense of deep quantum circuits. In order to build low-depth quantum algorithms that can be efficiently implemented on near-term quantum devices~\cite{preskill2018quantum}, the VQD algorithms based on McLachlan’s variational principle have been proposed for time evolution of a quantum state~\cite{LiBen17,EndSunLi20}. The accuracy of this hybrid quantum-classical simulation approach for computing the Green's function depends on the expressive power of the trial wave function~\cite{EndKurNak20,LibRizTac22,SakMizShi22}. To overcome this problem, an adaptive version of the variational QDS (AVQDS) approach has been proposed~\cite{yao2021adaptive,GyaLaw22,MooIadYao24}. The AVQDS method builds a compact wavefunction ansatz that is iteratively optimized for the specific problem, leading to highly accurate results with much simpler quantum circuits. These methods maintain a shallow circuit depth during the time evolution process by variationally minimizing the distance between a trial state that is prepared with parameterized circuits and the exact time-evolution state. One of the main limitations of variational approaches arises from the matrix inversion, which is sensitive to noise and condition number. Additionally, it is still a challenging task to extend variational QDS algorithms to long time simulations due to error accumulation from approximate ansatzes.

Cartan decomposition has been emerged as the state-of-the-art technique to construct an optimal decomposition of unitary operators. This method introduces an exact decomposition of the time evolution operator, which can be translated into fixed-depth circuits for arbitrarily long time evolution. Since the time evolution of a quantum system can be considered as a unitary operator, namely $e^{-i\mathcal{H}t}$, acting on a quantum state $\ket{\Phi_0}$, it is possible to reconstruct the time-evolution operator using Cartan decomposition of the Lie algebra generated by Pauli string terms in Hamiltonian. Recently, this technique has been employed to generate quantum circuits with fixed depth for the Hamiltonian simulation~\cite{KokSteWan22} and used to solve the Anderson impurity model in dynamical mean-field theory~\cite{steckmann2023mapping}. The latter work focuses on constructing optimal circuits via Cartan decomposition for solving the Fermi-Hubbard model. 

In this work, we present a detailed description of the QDS algorithm for computing the Green’s function based on Cartan decomposition. The exact ground states of the model systems are obtained using adaptive derivative-assembled pseudo-Trotter (ADAPT) VQE~\cite{grimsley2019adaptive}. As such, only time evolution of a single state is required to compute the retarded Green's function. A time-independent unitary transformation in a factorized form is employed so that analytic gradients can be formulated to perform Cartan decomposition efficiently. We apply this algorithm to simulate the Green's function in real time and extract the spectral functions for the two-site Fermi-Hubbard and transverse field Ising models, exhibiting high accuracy with respect to the exact numerical results.

\section{Method}\label{sec:method}
\subsection{Cartan Decomposition and $KHK$ Theorem}
Consider a $n$-qubits Hamiltonian 
\begin{equation}
    \mathcal{H} = \sum_i \lambda_i P_i,
\end{equation} 
where $P_i=\{I,X,Y,Z\}^{\otimes n}$ is an element of $S U\left(2^n\right)$ group and $X (Y, Z)$ is Pauli-X (Y, Z) operator. The closure (under commutation and linear combination) of a set of Pauli strings $\{P_i\}$ forms a Lie subalgebra $\mathfrak{g} \subset \mathfrak{s u}\left(2^n\right)$, which has a compact semi-simple Lie subgroup $\boldsymbol{G} \subset S U\left(2^n\right)$ in exponential map. A Cartan decomposition on Hamiltonian algebra $\mathfrak{g}$ is defined as finding an orthogonal split $\mathfrak{g}=\mathfrak{k} \oplus \mathfrak{m}$, satisfying
\begin{equation}
[\mathfrak{k}, \mathfrak{k}] \subset \mathfrak{k}, \quad[\mathfrak{m}, \mathfrak{m}] \subset \mathfrak{k}, \quad[\mathfrak{k}, \mathfrak{m}]=\mathfrak{m},   
\end{equation}
where $\mathfrak{k}$ is a maximal compact subalgebra of $\mathfrak{g}$. $\mathfrak{k}$ and $\mathfrak{m}$ is orthogonal under the killing form. This decomposition is labelled as $(\mathfrak{m}, \mathfrak{k})$. The Cartan subalgebra~\cite{hall2013lie} is defined as one of the maximal Abelian subalgebras of $\mathfrak{m}$, named $\mathfrak{h}$. 

In practical calculations, the Lie subalgebra is often partitioned by an involution. A Cartan involution is a Lie algebra homomorphism  $\theta: \mathfrak{g} \rightarrow \mathfrak{g}$, which satisfies $\theta(\theta(g))=g$ for any $g \in \mathfrak{g}$. This homomrophism preserves all commutators and distinguishes $\mathfrak{k}$ and $\mathfrak{m}$ by $\theta(\mathfrak{k})=\mathfrak{k}$ and $\theta(\mathfrak{m})=-\mathfrak{m}$.
Given a Cartan decomposition $\mathfrak{g}=\mathfrak{k} \oplus \mathfrak{m}$, for any element $m \in \mathfrak{m}$, there exists $K \in e^{\mathfrak{k}}$ and $h \in \mathfrak{h}$, such that~\cite{khaneja2001cartan,steckmann2023mapping}:
\begin{equation}
 m=K h K^{\dagger}.   
\end{equation}

Given a Cartan decomposition of $\mathfrak{g}(\mathcal{H})$ such that $\mathcal{H}\in\mathfrak{m}$, one can find an appropriate $K$ that gives $\mathcal{H}=K h K^\dagger$. In the Lie algebra, one can define a symmetric bilinear form as Killing form. For $\mathfrak{s u}\left(2^n\right)$, it is able to define the Killing form as (given $G_1, G_2 \in \mathfrak {g}$ viewed in their fundamental matrix representation):
\begin{equation}
    \kappa(G_1,G_2) = 2^{n+1} \mathrm{Tr}(G_1G_2).
\end{equation}
In this work, the trace part of standard Killing form is only chosen. The $K$ is determined by finding a local minimum of
\begin{equation}\label{eq:fk}
    f(K)=\kappa\left( K v K^{\dagger}, 
    \mathcal{H}\right),
\end{equation}
$v \in \mathfrak{h}$ is an element whose exponential map is dense in $e^{i \mathfrak{h}} $. For example, given $v=\sum_i \gamma^i h_i$, $h_i$ forms a basis for $\mathfrak{h}$, and $\gamma$ is an arbitrary transcendental number. At a local minimum $K_0$ of $f(K)$, 
\begin{equation}\label{eq:KHK}
    K_0^{\dagger} \mathcal{H} K_0=h \in \mathfrak{h},
\end{equation}
which determines $h$ and thus completes the decomposition.
The form of $K$ can be chosen as:
\begin{equation}
K= e^{\sum_i i a_i k_i} \quad \mathrm{or} \quad K=\prod_i e^{i a_i k_i},
\end{equation}
where $k_i$ is a Pauli string basis for $\mathfrak{k}$. Note that these two forms are equivalent due to compact Lie group and $\mathfrak{k}$ is closed under commutation. It is clear that finding $K$ is the computational bottleneck in the Cartan decomposition. Determining parameters of $K$ is a nonlinear optimization problem so that finding a minimum of Eq.~\eqref{eq:fk} is only possible for very small systems with few qubits due to the order of magnitude of $|\mathfrak{k}|$ is around $\frac{1}{2}|\mathfrak{g}|$~\cite{KokSteWan22}. 

In this work, we choose $K$ as:
\begin{equation}
K=\prod_i e^{i a_i k_i},\quad k_i \in \mathfrak{k}.
\end{equation}
To find a local minimum of the Killing form, we use the BFGS (Broyden-Fletcher-Goldfarb-Shanno) algorithm to optimize the parameters. The gradients required in BFGS can be obtained analytically. This comes from the structure of $K$, allowing us to apply $K$ on $v$ exactly. The $f(K)$ is written as:
\begin{equation}
   f(K)=\operatorname{Tr}\left(\prod_{i \uparrow} e^{i \theta_i k_i} v \prod_{i \downarrow} e^{-i \theta_i k_i} \mathcal{H} \right) 
\end{equation}
where $\uparrow(\downarrow)$ under the product means multiplication in an increasing (decreasing) order for $i$.To apply the exponentials of $k_i$, one can take advantage of the fact 
\begin{equation}
    e^{i \theta_i k_i}=\cos \theta_i I+i \sin \theta_i k_i .
\end{equation}
because $k_i^2=I$ for $k_i$ being an Pauli operators.
After applying all exponentials to $v$,
\begin{equation}
   f(K)=\operatorname{Tr}\left(m_0 \mathcal{H}\right) 
\end{equation}
is obtained. Here,
\begin{equation}
m_0=\prod_{i \uparrow} e^{i \theta_i k_i} v \prod_{i \downarrow} e^{-i \theta_i k_i} \in \mathfrak{m}
\end{equation}
Up to this point, in the worst case only $\mathcal{O}(|\mathfrak{k}||\mathfrak{m}|)$ operations are performed. $|\mathfrak{k}|$ exponentials are applied to an element of $\mathfrak{m}$, which has at most $|\mathfrak{m}|$ Pauli terms. Multiplying $m_0$ and $\mathcal{H}$ requires $\mathcal{O}(|\mathfrak{m}|)$ calculations and can be neglected in the $|\mathfrak{k}| \gg 1$ limit corresponding to large system size limit, which leads to $\mathcal{O}(|\mathfrak{k}||\mathfrak{m}|)$ time complexity to calculate $f(K)$.
The gradient of $K$ can be expressed as
\begin{equation}
    \frac{\partial K}{\partial \theta_j}=\prod_{i<j, \uparrow} e^{i \theta_i k_i} i k_j \prod_{i \geq j, \uparrow} e^{i \theta_i k_i},
\end{equation}
leading to the analytical expression for the gradient of the function:
\begin{equation}
\begin{aligned}
&\frac{\partial f(K)}{\partial \theta_j} =i\operatorname{Tr}\left(\prod_{i<j, \uparrow} e^{i \theta_i k_i} k_j \prod_{i \geq j, \uparrow} e^{i \theta_i k_i} v \prod_{i \downarrow} e^{-i \theta_i k_i} \mathcal{H}\right) \\
&-i\operatorname{Tr}\left(\prod_{i \uparrow} e^{i \theta_i k_i} v \prod_{i \geq j, \downarrow} e^{i \theta_i k_i} k_j \prod_{i<j, \downarrow} e^{-i \theta_i k_i} \mathcal{H}\right).
\end{aligned}
\end{equation}
Using the permutation property of the trace function, we can calculate the each term applied on $v$ and $\mathcal{H}$. Denoting
\begin{equation}
    v_j = \prod_{i \geq j, \uparrow} e^{i \theta_i k_i} v \prod_{i \geq j, \downarrow} e^{-i \theta_i k_i},
\end{equation}
and
\begin{equation}
    \mathcal{H}_j = \prod_{i<j, \downarrow} e^{-i \theta_i k_i} \mathcal{H}\prod_{i<j, \uparrow} e^{i \theta_i k_i},
\end{equation}
the gradient of the function can be written as:
\begin{equation}
    \frac{\partial f(K)}{\partial \theta_j} = i\mathrm{Tr}(k_jv_j\mathcal{H}_j)-i\mathrm{Tr}(v_jk_j\mathcal{H}_j).
\end{equation}
Consequently, one can calculate the analytical gradient of each parameter in parallel. Finally, eq.(\ref{eq:KHK}) leads to the desired unitary for time evolution:
\begin{equation}\label{eq:tmop}
    U(t) = K_0e^{-iht}K_0^\dagger.
\end{equation}
Since $\mathfrak{h}$ is Abelian and $h\in\mathfrak{h}$, the $e^{-iht}$ is equal to $\prod_{i=0}^N e^{-i\lambda_ih_it}$ with $h=\sum_{i=0}^N \lambda_i h_i$. Every term in $U(t)$ is written as $e^{-i\lambda_i P_i}$ and therefore a quantum circuit for $e^{-i\lambda_i P_i}$ can easily be constructed. 

\subsection{Green's Function}
Given a fermionic system, the Hamiltonian $\mathcal{H}$ consists of fermionic creation ($\hat{c}^\dagger_{a}$) and annihilation  ($\hat{c}_{a}$) operators. The subscript $a$ denotes a general index labeling the fermionic modes, which, in electronic structure calculations, typically corresponds to a single-particle orbital in a chosen basis.
The corresponding retarded Green's function at zero temperature is:
\begin{equation}\label{eq:rGF}
    G^R_{ab}(t)=-i\Theta(t)\braket{\hat{c}_{a}(t)\hat{c}^\dagger_{b}(0)+\hat{c}^\dagger_{b}(0)\hat{c}_{a}(t)}_0
\end{equation}
which $\Theta(t)$ is Heaviside step function and $\braket{\dots}_{0} = \braket{\Psi|\dots|\Psi}$ denotes the expectation value by the ground state $\ket{\Psi}$ of the Hamiltonian. $\hat{c}^\dagger_{a}(t)$ and $\hat{c}_{a}(t)$, defined as 
\begin{equation}
\begin{split}
        \hat{c}^\dagger_{a}(t) &= e^{i\mathcal{H}t}
        \hat{c}^\dagger_{a}e^{-i\mathcal{H}t} \\ 
        \hat{c}_{a}(t) &= e^{i\mathcal{H}t} \hat{c}_{a}e^{-i\mathcal{H}t},
\end{split}
\end{equation}
denote the Heisenberg representations of the fermionic creation and annihilation operators $\hat{c}^\dagger_{a}$ and $\hat{c}_{a}$, respectively.

By the ground-state wave function $\ket{\Psi}$ is the eigenstate of the Hamiltonian $\mathcal{H}$ with the eigenvalue of $E_g$, one obtains
\begin{equation}
    e^{-iHt}\ket{\Psi} = e^{-iE_gt}\ket{\Psi}. 
\end{equation}
Consequently, the retarded Green's function at $t>0$ can be further simplified as
\begin{equation}\label{eq:GFrat}
\begin{aligned}
    G^R_{ab}(t)=&-ie^{iE_gt}\braket{\hat{c}_{a}e^{-i\mathcal{H}t}\hat{c}^\dagger_{b}}_0& \\
    &-ie^{-iE_gt}\braket{\hat{c}^\dagger_{b}e^{i\mathcal{H}t}\hat{c}_{a}}_0.&
\end{aligned}
\end{equation}
So the problem reduces to finding a way to compute the inner product on quantum computer. After Jordan-Wigner transformation~\cite{jordan1928ber}:
\begin{equation}
\begin{aligned}
    \hat{c}^\dagger_{a} &= Q^\dagger_a \otimes Z_{a-1}\otimes Z_{a-2} \dots \otimes Z_{1}\\
    \hat{c}_{a} &= Q_a \otimes Z_{a-1}\otimes Z_{a-2} \dots \otimes Z_{1}\ , 
\end{aligned}
\end{equation}
which $Q^\dagger_a = \frac{1}{2}(X_a-iY_a)$ and $Q_a = \frac{1}{2}(X_a+iY_a)$. Consequently,
\begin{equation}
    \braket{\hat{c}_{a}e^{-i\mathcal{H}t}\hat{c}^\dagger_{b}}_0 = 
    \sum_{ij}\braket{\Psi|P_{i} e^{-i\mathcal{H}t} P_{j}|\Psi}.
\end{equation}
The state $Q_a\ket{\Psi}$ can be directly prepared in real quantum computer devices and the time evolution of these states can be directly evolved through the quantum circuits obtained by the previous method:
\begin{equation}\label{initial-state}
      U(t)Q_a\ket{\Psi} = K_0e^{-iht}K_0^\dagger Q_a\ket{\Psi}.
\end{equation}
After obtaining the simulated state at the corresponding time, we can use the Hadamard test~\cite{aharonov2006polynomial} method to obtain the inner product of this state and other states. 

For simplicity, we consider the Green’s function $G^R_{ab}(t)$ as $a=b=(k,\sigma)$ throughout this work, which means the Green’s
function in the momentum space with identical spin. The spectral function $A_{k,\sigma}(\omega)$ can be defined by the Fourier transform of the Green’s function:
\begin{equation}
\begin{aligned}
\tilde{G}_{k,\sigma}^{R}(\omega) & =\int_{-\infty}^{\infty} d t e^{i(\omega+i \eta) t} G_{k,\sigma}^{R}(t) \\
& =: \int_{-\infty}^{\infty} d \omega^{\prime} \frac{A_{k,\sigma}\left(\omega^{\prime}\right)}{\omega-\omega^{\prime}+i \eta}
\end{aligned}
\end{equation}
where $\eta\rightarrow 0^{+}$ is a factor to assure the convergence of the integral. The spectral function $A_{k,\sigma}(\omega)$ and the Green’s function $\tilde{G}_{k,\sigma}^{R}(\omega)$ have a relation:
\begin{equation}\label{eq:spectral-fun}
    A_{k,\sigma}(\omega)=-\pi^{-1} \operatorname{Im} \tilde{G}_{k,\sigma}^{R}(\omega) .
\end{equation}

\section{Numerical Demonstration}
\subsection{Two-Site Fermi-Hubbard Model}
We first apply the Cartan decomposition-based method to calculate the Green’s function and the spectral function at zero temperature of Fermi-Hubbard model. The Hamiltonian of two sites Fermi-Hubbard model with the particle-hole symmetry is:
\begin{equation}\label{eq:Fermi-Hubbard}
    \mathcal{H} = t\sum_{\sigma=\uparrow,\downarrow}(\hat{a}_{0,\sigma}^\dagger\hat{a}_{1,\sigma}+h.c.) + U\sum_{i = 0}^1(\hat{n}_{i,\uparrow}-\frac{1}{2})(\hat{n}_{i,\downarrow}-\frac{1}{2}),
\end{equation}
where $t$ characterizes hopping amplitude between nearest-neighbor sites. In the following calculations, $t$ is set to be -1. $U$ is the on-site interaction energy between fermions of opposite spin occupying the same lattice site~\cite{gutzwiller1963effect,kanamori1963electron,royal1869proceedings}. The local density operator $\hat{n}_{i,\sigma} = \hat{a}_{i,\sigma}^\dagger\hat{a}_{i,\sigma}$.
The creation and annihilation operators in real space and momentum space are related through the following transformation
\begin{equation}\label{eq:momentum space op1}
    \hat{c}^\dagger_{k,\sigma} = \frac{1}{\sqrt{N}}\sum_{j} \hat{a}^\dagger_{j,\sigma} e^{-ikj},
\end{equation}
\begin{equation}\label{eq:momentum space op2}
    \hat{c}_{k,\sigma} = \frac{1}{\sqrt{N}}\sum_{j} \hat{a}_{j,\sigma}e^{ikj},
\end{equation}
where $N$ indicates the number of sites. As a result, the retarded Green’s function in momentum space is defined as
\begin{equation}
    G^R_{k,\sigma}(t) = \frac{1}{N}\sum_{i,j}G^R_{ij,\sigma}(t)e^{ik(i-j)}.
\end{equation}
In the following, we choose $k=0$ and $\pi$ to assess the performance of the Cartan decomposition-based method. In the case of $k=0$, the Eq.~\eqref{eq:momentum space op1} and Eq.~\eqref{eq:momentum space op2} are:
\begin{equation}
    \hat{c}^\dagger_{0,\sigma} = 
    \frac{1}{\sqrt{2}}
    \left( \hat{a}^\dagger_{0,\sigma} + \hat{a}^\dagger_{1,\sigma}
    \right),
\end{equation}
\begin{equation}
    \hat{c}_{0,\sigma} = \frac{1}{\sqrt{2}}
    \left( \hat{a}_{0,\sigma} + \hat{a}_{1,\sigma}
    \right).
\end{equation}
And in the case of $k=\pi$, they are
\begin{equation}
    \hat{c}^\dagger_{\pi,\sigma} = 
    \frac{1}{\sqrt{2}}
    \left( \hat{a}^\dagger_{0,\sigma} - \hat{a}^\dagger_{1,\sigma}
    \right),
\end{equation}
\begin{equation}
    \hat{c}_{\pi,\sigma} = \frac{1}{\sqrt{2}}
    \left( \hat{a}_{0,\sigma} - \hat{a}_{1,\sigma}
    \right).
\end{equation}

Here, each site is related to two spin orbits and  each spin orbit is mapped onto a qubit. As a result, the site index corresponds to the qubit index by $(i,\uparrow)\rightarrow(2i)$ and $(i,\downarrow)\rightarrow(2i+1)$. After Jordan-Wigner transformation by using the library OpenFermion ~\cite{mcclean2020openfermion}, the Fermi-Hubbard Hamiltonian of Eq.~\eqref{eq:Fermi-Hubbard} is written as:
\begin{equation}
\begin{aligned}
    \mathcal{H}_q =& \frac{t}{2}(X_0Z_1X_2 + Y_0 Z_1Y_2 + X_1 Z_2 X_3 + Y_1 Z_2 Y_3) \\
    &+ \frac{U}{4}(Z_0 Z_1 + Z_2 Z_3).  
\end{aligned}
\end{equation}
The subscript $i$ indicates the Pauli operator acts on $i$-th qubit. Each term in the qubit Hamiltonian is considered as a basis element in the Hamiltonian Lie algebra. The dimension of the total Hamiltonian Lie algebra $\mathfrak{g}$ is 24, and $\mathfrak{k}$ and $\mathfrak{h}$ have 8 basis elements, respectively. They are consisted by following elements:
\begin{equation}
\begin{aligned}
\mathfrak{k} = \mathbf{span}
\{&X_0Y_2,Y_0X_2,Y_1X_3,X_1Y_3,\\
       &Y_0Z_1X_2Z_3,X_0Z_1Y_2Z_3,\\
       &Z_0X_1Z_2Y_3,Z_0Y_1Z_2X_0\},
\end{aligned}
\end{equation}
\begin{equation}
\begin{aligned}
\mathfrak{h} = \mathbf{span}
\{ &Y_0Z_1Y_2,X_0Z_1X_2,Y_1Z_2Y_3,\\
   &X_1Z_2X_3,X_0X_2Z_3,Y_0Y_2Z_3,\\
   &Z_0X_1X_3,Z_0Y_1Y_3
\}.
\end{aligned}
\end{equation}

\begin{figure*}[ht]
    \centering
    \includegraphics[width=0.8\linewidth]{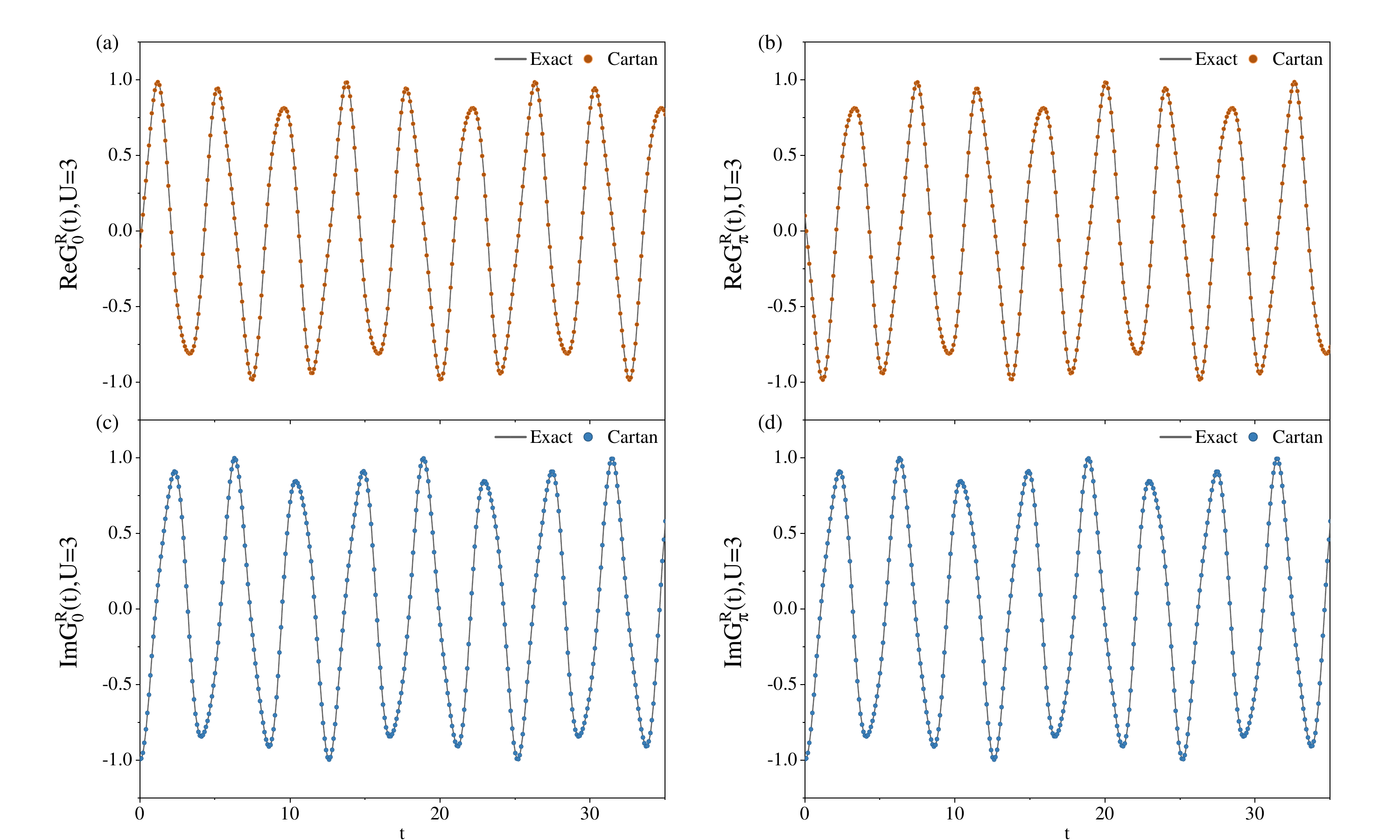}
    \caption{Numerical simulation of the exact method and  Cartan algorithm to compute the Green’s function in real time $G_k^R(t)$ for the model of $U=3$. The dynamics in real time is calculated from $t$ = 0 to $t$ = 35 with step $dt$ = 0.1.}
    \label{fig:GF-t-U3}
\end{figure*}

\begin{figure*}[!htbp]
    \centering
    \includegraphics[width=0.8\linewidth]{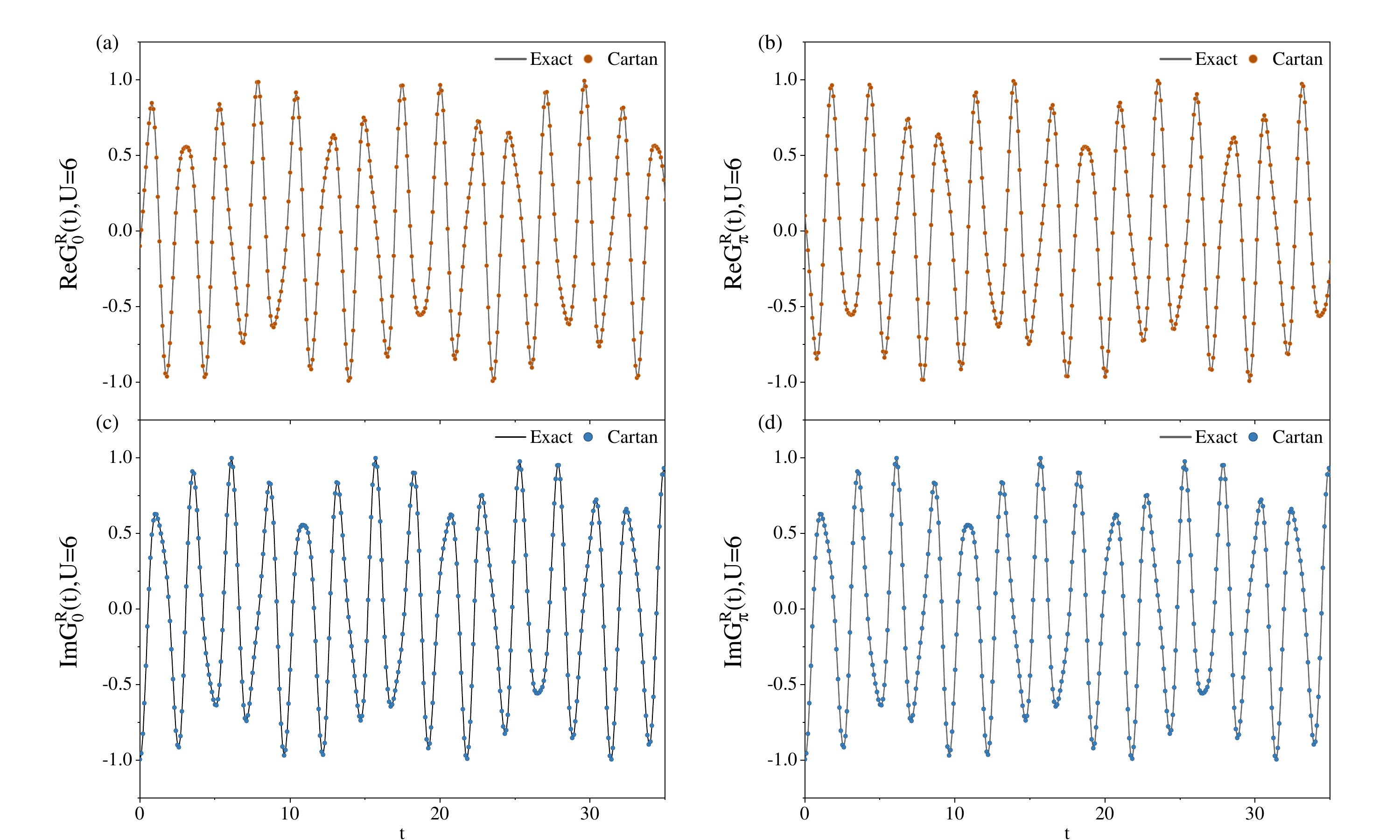}
    \caption{Numerical simulation of the exact method and Cartan algorithm to compute the Green’s function in real time $G_k^R(t)$ for the model of $U=6$. The dynamics in real time is calculated from $t$ = 0 to $t$ = 35 with step $dt$ = 0.1.}
    \label{fig:GF-t-U6}
\end{figure*}

Due to time reversal symmetry, only the case of $(k,\uparrow)$ is calculated here and this index is abbreviated as $k$. We present a detailed study of the retarded Green's function $G^R_k(t)$ and corresponding spectral function $A_k(\omega)$ for two representative interaction strengths, $U=3$ and $U=6$. Fig.~\ref{fig:GF-t-U3} and Fig.~\ref{fig:GF-t-U6} show the numerical simulation result of Green's functions in real time. The Cartan approach is able to reproduce the exact results over total time, indicating that the simulation process has high fidelity. By analyzing both real-time dynamics and frequency-resolved spectral features at momentum points $k=0$ and $k=\pi$, we identify clear signatures of the crossover from a strongly correlated metallic phase to a Mott insulating~\cite{mott1968metal} regime. Our results reveal the interplay between temporal coherence, quasiparticle lifetime, and the emergence of Hubbard bands with increasing interaction strength.

Like any complex function, a retarded Green's function has both a real and an imaginary part. $(a),(b)$ in Fig.~\ref{fig:GF-t-U3} and Fig.~\ref{fig:GF-t-U6} present the real parts of $G^R_k(t)$, and $(c),(d)$ present the imaginary parts. At $U=3$, the real part of $G^R_k(t)$ displays pronounced oscillations with relatively slow decay. The imaginary part remains finite and oscillatory over long timescales. These features indicate the presence of well-defined, long-lived quasiparticles and sustained coherence in the electron propagation. But at $U=6$, both the real and imaginary components decay rapidly with time. The oscillatory behavior observed at 
$U=3$ is strongly suppressed, signifying a loss of temporal coherence. The fast decay reflects reduced quasiparticle lifetime and enhanced localization due to strong electron repulsion. These observations illustrate the dynamical signatures of the Mott transition, wherein increasing interaction strength leads to the breakdown of quasiparticle excitations and electronic delocalization.

\begin{figure*}[ht]
    \centering
    \includegraphics[width=0.8\linewidth]{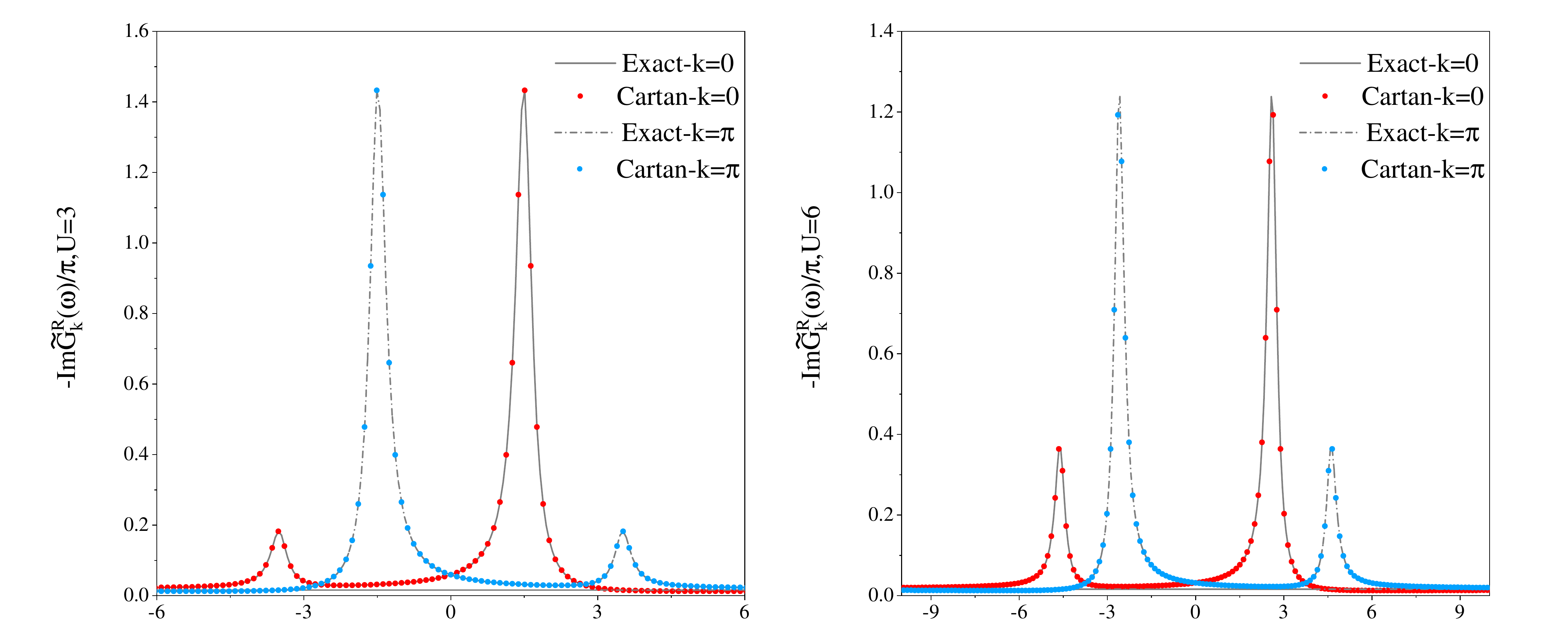}
    \caption{Numerical simulation of the spectral function. We take $\eta$ = 0.2 for the calculation of the spectral functions from eq.(\ref{eq:spectral-fun})}
    \label{fig:GF-w}
\end{figure*}

Fig.~\ref{fig:GF-w} shows the spectral functions $A_k(\omega)$ calculated from retarded Green's functions $G_k^R(t)$ obtained by different methods. The spectral functions $A_k(\omega)$ provide direct access to the single-particle excitation spectrum. These results further support the time-domain findings. The spectral function exhibits a clear two-peak structure, corresponding to the lower and upper Hubbard bands at $U=6$. This redistribution of spectral weight away from the Fermi level signifies the breakdown of coherent quasiparticles and the onset of insulating behavior.

The comparison between $k=0$ and $k=\pi$ also reveals key differences in how momentum influences the spectral and dynamical properties. As $U$ increases, the spectral peaks shift away from the Fermi level, and the effective bandwidth is strongly renormalized. The system transitions from a dispersive metallic band to a two-band structure separated by an interaction-induced gap.

The numerical results provide a comprehensive view of the interaction-driven evolution in the 1D Hubbard model, based on exact time-domain and frequency-domain data. At intermediate coupling $U=3$, the system behaves as a correlated metal, characterized by oscillatory $G_k^R(t)$, long quasiparticle lifetimes, and finite spectral weight at the Fermi level. But at strong coupling $U=6$, the system undergoes a Mott transition, featuring fast decay of $G_k^R(t)$, loss of coherence, vanishing spectral weight at $\omega=0$, and the emergence of upper and lower Hubbard bands. These findings confirm the consistency between time-domain and frequency-domain characterizations of the metal-insulator transition~\cite{mott1968metal,meinders1993spectral}.

\subsection{One-dimensional Spin Chains}
In this section, we apply Cartan method to study many-body physical phenomena in transverse field Ising model (TFIM)~\cite{radicevic2018spin}. Consider the Hamiltonian of TFIM with open boundary conditions:
\begin{equation}
    H_{\textbf{TFIM}} = \sum_{i=0}^{n-1} Z_{i+1}Z_i + h_x\sum_{i=0}^n X_i.
\end{equation}
The corresponding Hamiltonian Lie algebra is generated by the set:
\begin{equation}
    \mathbf{G}_{\textbf{TFIM}} = \left\{ Z_{i+1}Z_i ,X_i | 0\leq i \leq n\right\},
\end{equation}
The total Hamiltonian Lie algebra $\mathfrak{g}$ can be written as:
\begin{equation}
    \mathfrak{g} = \mathbf{span}\{X_i,\widehat{Z_iZ_j}, \widehat{Y_iZ_j},\widehat{Z_iY_j},\widehat{Y_iY_j}~|~0\leq i< j \leq n\},
\end{equation}
the $\mathfrak{k}$ and $\mathfrak{h}$ is:
\begin{equation}
\begin{aligned}
\mathfrak{k} = \mathbf{span}
\{\widehat{Y_iZ_j},\widehat{Z_iY_j}~|~0\leq i< j \leq n\},
\end{aligned}
\end{equation}
\begin{equation}
\begin{aligned}
\mathfrak{h} = \mathbf{span}
\{ &Z_0Z_1,Z_1Z_2,\dots,Z_{n-1}Z_n,\\
   &Y_0X_1\dots X_{n-1}Y_n
\}.
\end{aligned}
\end{equation}
which we use $\widehat{P_iP_j}$ to represent $P_iX_{i+1}\dots X_{j-1}P_j$ for $P=Y$ or $Z$.
We define the Lieb-Robinson type~\cite{mcdonough2025lieb} retarded correlation function by:
\begin{equation}
    G^R_k(t)= \frac{1}{N_k}\sum_{i=0}^{N_k}-i\Theta(t)\braket{[Z_i(t),Z_0]}_0,
\end{equation}
which different $k$ means different sizes and We choose $N_1 = 2, N_2 = 4, N_3 = 6$. Consistent with the previous definition, $\tilde{G}_{k}^{R}(\omega)$ is the Fourier transform of the retarded Green’s function. We investigate the behavior of the spectral function $A_k(\omega)$ as a function of frequency $\omega$ for systems of different sizes.

\begin{figure}[!htbp]
    \centering
    \includegraphics[width=0.9\linewidth]{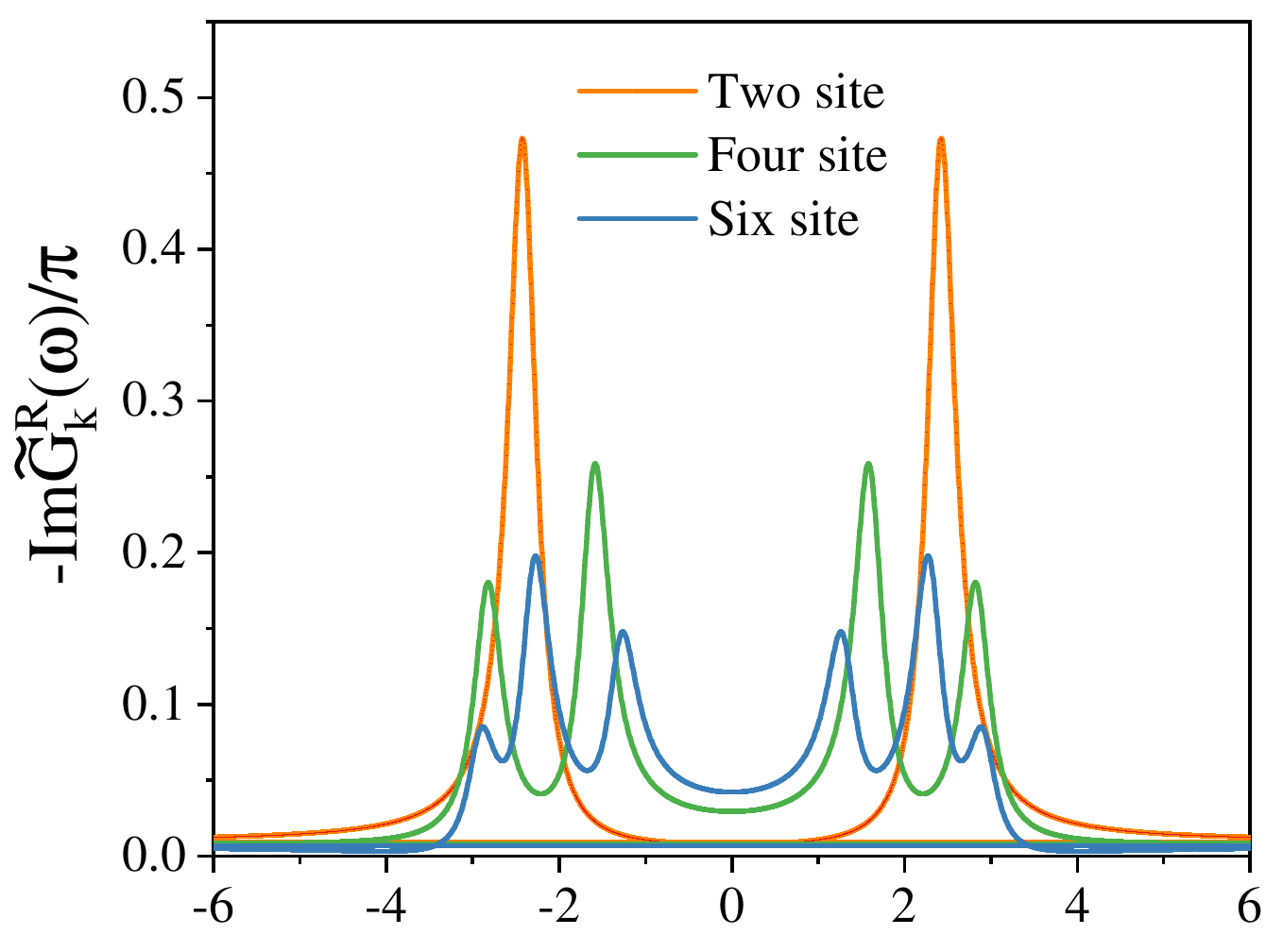}
    \caption{Numerical simulation of the spectral function for different spin chains. We take $\eta$ = 0.2 for the calculation of the spectral functions}
    \label{fig:Ising-specral}
\end{figure}

Fig.~\ref{fig:Ising-specral} shows the spectral functions $A_k(\omega)$ calculated from retarded Green's functions. The three curves in the figure correspond to two-site, four-site, and six-site systems, and demonstrate how the spectral function evolves as the system size increases. For the two-site system (orange curve), the spectral function exhibits sharp peaks, indicating that in smaller systems, quantum effects are more pronounced. As the system size increases to four sites and six sites, the main peaks broaden, and secondary peaks emerge, reflecting the increased complexity of the system due to more interactions between the particles.

This evolution suggests that smaller systems are dominated by local effects, while larger systems exhibit more complex and continuous spectral distributions, reflecting the emergence of global effects. Furthermore, as the system size grows, the peaks in the spectral function shift toward lower frequencies 
$(\omega=0)$ and become broader, suggesting a transition to a more stable state. This behavior is likely linked to the quantum phase transition characteristics in the transverse field Ising model. In the six-site system, the spectral function becomes smoother, with diminishing differences between peaks, hinting at the approach of a quantum critical region where the system exhibits universal behavior.

The broadening of peaks and the appearance of additional features in the spectral function as the system size increases point to the system’s tendency toward a more stable, collective behavior, which is a hallmark of quantum criticality. This transition from sharp, localized features to broader, more continuous distributions suggests that, in larger systems, the system is less sensitive to local perturbations and more influenced by long-range interactions. Such trends are indicative of the onset of quantum phase transitions, where the system undergoes qualitative changes in its behavior, often linked to critical points that separate distinct phases.

As the system size increases, the spectral function evolves from sharp, localized features in small systems to more complex, smooth distributions in larger systems, reflecting the transition from local to global behavior. This shift provides crucial insights into the quantum phase transitions in the TFIM, especially in the large-system limit. The observed changes in the spectral function across different system sizes underscore the critical role that system size plays in the manifestation of quantum effects and phase transitions, contributing to a deeper understanding of quantum critical phenomena in condensed matter physics. The results suggest that, in the large system limit, the model's behavior becomes increasingly universal, offering valuable information for exploring quantum criticality in a variety of systems.

\section{Conclusion}
In this work, we have presented an efficient algorithm for computing the single-particle Green’s function on near-term quantum computers by leveraging Cartan decomposition of unitary time evolution operators. By reformulating quantum dynamics as a Cartan-structured unitary transformation, we enable real-time simulation of Green's functions with fixed-depth quantum circuits, independent of the total simulation time. This significantly reduces circuit complexity and makes the method highly suitable for implementation on near-term quantum devices. To enhance numerical efficiency, we introduced analytical gradients for the decomposition based on a factorized form, allowing fast and accurate optimization using classical resources. The method is applied to two prototypical quantum many-body models-the two-site Fermi-Hubbard model and the transverse-field Ising model-demonstrating excellent agreement with exact results across a range of interaction strengths and system sizes.

The simulation results not only reproduce spectral functions with high fidelity but also capture key physical phenomena, such as the Mott metal-insulator transition and finite-size scaling behavior near quantum criticality. These findings confirm that the Cartan decomposition-based approach is capable of preserving essential dynamical and spectral information while maintaining compatibility with current quantum hardware constraints. We believe this method paves the way for hardware-efficient quantum simulations of dynamical correlation functions in condensed matter physics, quantum chemistry, and materials science. It also lays a foundation for incorporating Cartan decomposition into broader classes of quantum algorithms for time evolution, response functions, and spectral analysis in the NISQ era. 

The dimension of $K$ introduced in this work scales exponentially as the system size for simulating the Heisenberg model and molecules. Consequently, Cartan decomposition is limited to small-sized simulations up to ~10 qubits. In principle, we can optimize the form of K to improve the scalability by exploiting the intrinsic locality of the quantum systems. For example, we can utilize Zassenhaus formula to approximate the time-evolution operator and apply Cartan decomposition to the block Hamiltonian algebra to reduce the computational scaling. 


\section{Acknowledgments}
This work was supported by Innovation Program for Quantum Science and Technology (2021ZD0303306), the National Natural Science Foundation of China (22393913, 22422304), Anhui Initiative in Quantum Information Technologies (AHY090400).

\bibliography{qc}
\bibliographystyle{rsc}
\end{document}